\documentclass[prb,twocolumn,superscriptaddress,floatfix,showkeys]{revtex4}
\usepackage{graphicx,amssymb,bm,natbib,amsfonts,amsmath,graphics,color}
\usepackage[bookmarks=false]{hyperref}

\begin{document}

\title{Mapping the spin-dependent electron reflectivity of Fe and Co ferromagnetic thin films}

\author{J. Graf} \email{jeff.graf@a3.epfl.ch} \affiliation{Materials Sciences Division, Lawrence Berkeley National Laboratory, Berkeley, California 94720}
\author{C. Jozwiak} \affiliation{Dept. of Physics, University of California Berkeley, California 94720}
\author{A.K. Schmid} \affiliation{Materials Sciences Division, Lawrence Berkeley National Laboratory, Berkeley, California 94720}
\author{Z. Hussain} \affiliation{Advanced Light Source, Lawrence Berkeley National Laboratory, Berkeley, California 94720} 
\author{A. Lanzara} \affiliation{Materials Sciences Division, Lawrence Berkeley National Laboratory, Berkeley, California 94720} \affiliation{Dept. of Physics, University of California Berkeley, California 94720}

\date{\today}

\begin{abstract} Spin Polarized Low Energy Electron Microscopy is used as
a spin dependent spectroscopic probe to study the spin dependent specular
reflection of a polarized electron beam from two different magnetic thin
film systems: Fe/W(110) and Co/W(110). The reflectivity and spin-dependent
exchange-scattering asymmetry are studied as a function of electron
kinetic energy and film thickness, as well as the time dependence. The
largest value of the figure of merit for spin polarimetry is observed for
a 5 monolayer thick film of Co/W(110) at an electron kinetic energy of
2eV.  This value is 2 orders of magnitude higher than previously obtained
with state of the art Mini-Mott polarimeter. We discuss implications of
our results for the development of an electron-spin-polarimeter using
the exchange-interaction at low energy.  \end{abstract}

\keywords{exchange scattering; Co/W(110); Fe/W(110); ferromagnetic thin film; electron spin polarimeter; figure of merit; SPLEEM; Sherman function}

\maketitle

\section{INTRODUCTION}
\label{sec:INTRODUCTION}

State of the art electron spectrometers have reached an energy and momentum resolution that allows probing subtle and complex many-body effects. However the low efficiency of spin-detection impedes experimental
progress towards our improved understanding of the spin degree of freedom in complex and magnetic materials.  In terms of spin-detector development,
a quantity of great interest is defined as the Figure of Merit (FOM) and is proportional to the inverse square of the statistical error in an electron counting experiment to measure the polarization of an incident
beam \cite{kessler:1985}. Conventional spin-polarimeters (mini-Mott detectors) have a FOM below 2x10$^{-4}$ ~\cite{pierce:550}. This low FOM has long been recognized as problematic. For well over one decade, various groups have worked on the idea to use exchange-scattering of low energy electrons by ferromagnetic surfaces as a promising route toward achieving significantly improved efficiency. For example, detailed investigations of Fe(110) surface have shown that FOM of the order of 8x10$^{-3}$ can be achieved ~\cite{Fahsold:541,Hammond:6131}.
Along this direction, different surfaces have been investigated to increase the FOM, the lifetime and the reproducibility~\cite{hillbrecht:1229,jungblut:615}. Recently, FOM up to 6x10$^{-3}$ have been achieved with a robust system, Fe(001)-p(1x1)O~\cite{bertacco:265,bertacco:2050}, and a spin resolved electron spectrometer has been built ~\cite{bertacco:3867}. A different approach suggested even higher FOM (up to  5x10$^{-2}$) using high quality ultra-thin Co or Fe films on W(110)~\cite{zdyb:1485,zdyb:166403}. Unfortunately, the microscopic size of this ultra-thin film is problematical in spin polarimetry applications~\cite{bertacco:3867}. Here, we demonstrate that a larger film of lower quality that can be easily grown and reproduced at room temperature can indeed reliably achieve very high FOM.

In this paper, we report a detailed study of the reflectivity and spin-dependent exchange-scattering asymmetry as a
function of electron kinetic energy and film thickness and as a function
of time for two different magnetic thin film systems grown at room temperature: Fe/W(110) and
Co/W(110). The results confirm the possibility of constructing novel
spin polarimeters with FOM up to two orders of magnitude higher than state of the art mini-Mott polarimeter and stray fields significantly lower than current exchange-scattering-based polarimeters using bulk ferromagnets.

\section{EXPERIMENTAL TECHNIQUE}

The data reported here are taken with a Spin Polarized Low Energy
Electron Microscope (SPLEEM)\cite{bauer:2327} located in the National
Center for Electron Microscopy of the Lawrence Berkeley National Laboratory. SPLEEM is a powerful tool to
image the dynamics of surface magnetic microstructures in the sub-second
timescale, and its sensitivity to the sample surface allows the imaging
of atomic steps with a lateral resolution of $\sim$20nm. Here, we use
it as a fast and powerful spin dependent electronic probe by recording
the number of elastically backscattered electrons as a function of their
spin and energy.

Cobalt and iron films were prepared at room temperature by electron beam
evaporation of high-purity material on a (110) tungsten crystal. The
tungsten substrate was cleaned by successive flash heating cycles in an
O$_2$ environment of 10$^{-8}$ Torr, followed by flash heating in UHV up
to $\approx$2000$^{\circ}$C.  The substrate cleanliness was monitored
by Auger electron spectroscopy and no traces of contamination were
detected. The base pressure was on the order of 10$^{-11}$ Torr and was
in the low 10$^{-10}$ Torr region during film deposition. The deposition
rate was measured by direct observation with SPLEEM and adjusted to
approximately 0.1 monolayer (ML) per minute. Both Co and Fe grow in a
quasi layer-by-layer mode at room temperature as previously reported
\cite{gradmann:539,gardiner:213}. More specifically, Fe grows in a
pseudomorphic mode for the first 1.8 ML, and then relaxes to a bulk bcc
structure. The cobalt film adopts pseudomorphic structure for the first
ML and then transforms to a close-packed structure \cite{bauer:9365}.

\begin{figure} \includegraphics[width=8.5cm]{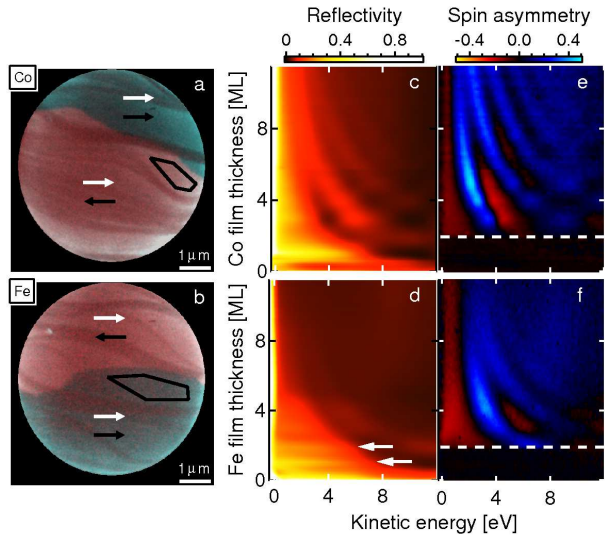}
\caption{\label{fig:1}(Color online) Panels (a) and (b) show SPLEEM
images of Co/W(110) and Fe/W(110) films, respectively. For both films
the magnetic easy axis points along the horizontal direction in the
images. Both images were recorded with electron spin polarization aligned
to the right (white arrow). The magnetic domains are shown with blue
and red contrast, magnetization direction of the domains is indicated
by black arrows. Black contours are the regions of interest for average
reflectivity measurements. (c,d) Color-scale plots of the measured
reflectivity as a function of the electron kinetic energy (horizontal
axis) and film thickness (vertical axis) for both thin-film systems. (e,f)
Color-scale plots of the spin asymmetry of the reflectivity.} \end{figure}

Figures \ref{fig:1}a and b show images of the magnetic surface
for Co/W(100) (1a) and Fe/W(110) (1b) at a certain energy and film
thickness. The bright areas represent flat terraces on the crystal
surface and the darker curving lines and bands represent the monoatomic
steps and multilayer step bands of the W(110) surface.  A blue and a red
region can be observed for both images. These two regions represent two
magnetic domains with opposite magnetization (black arrows). The white
arrows represent the majority spin orientation of the illumination
electron beam. Within the imaged field of view, regions of interest
(ROI) where substrate step-density is low and film magnetization is
homogeneously aligned in a particular magnetic domain (see black contour
in figure \ref{fig:1}a and b) are selected. The average intensity at a
given electron energy, film thickness and spin orientation is extracted
in these regions. The spin-polarization of the illuminating electron
beam was adjusted to be aligned with the magnetic easy axis of the
film. Continuous sequences of images of the surfaces were then recorded
automatically, during the simultaneous deposition of the ferromagnetic
thin films. For each set of two consecutive images, the spin orientation
of the electron beam was toggled to be parallel and anti-parallel to the
film magnetization. The deposition rate was adjusted to be sufficiently slow
so that during the deposition of each ML we were able to ramp several
times the electron beam energy from 0 to 13 eV in steps of 0.2 eV (with
respect to the vacuum level). The average intensity in the selected
ROI is then measured for each image and normalized with respect to the
electron illumination intensity.

\section{RESULTS}
\label{sec:RESULTS}

Figure \ref{fig:1}c (d) shows a color plot representing the average
reflectivity for cobalt (iron) versus the electron energy (horizontal
axis) and the film thickness (vertical axis). The variations in the
reflectivity seem to have at least two periodic components. The component
that is visibly energy independent (see white arrows in fig. \ref{fig:1}d)
is due to the varying density of steps during deposition. The reflectivity
minima (in black) coincide with half completed monolayers (higher
surface step density) and the maxima (in red) occur with the completion
of each monolayer (low surface step density). These growth oscillations
are most clear at small film thickness and are used to calibrate the
film deposition rate. The interesting component is energy dependent
and looks like hyperbolic shaped maxima (in red) and minima (in black)
in the reflectivity plots (more pronounced for Co/W(100)). This is a
bulk and intrinsic effect known as the ``Quantum Size Effect'' (QSE)
\cite{zdyb:1485,scheunemann:787}.  More specifically, a ferromagnetic slab
acts like a resonant cavity and induces interference patterns. The QSE
is more pronounced for Co/W(110) (fig. \ref{fig:1}c) than for Fe/W(110)
(fig. \ref{fig:1}d) films. We attribute this attenuation to previously
reported kinetic roughening during room temperature growth of Fe/W(110)
\cite{zdyb:1485}.

In figure \ref{fig:1}e and f, we report the spin asymmetry
versus energy and film thickness for Co/W(110) and Fe/W(110) films
respectively. Here, the spin asymmetry (or Sherman function) is defined by
\[A_s=\frac{1}{P}\frac{R_{\uparrow\downarrow}-R_{\uparrow\uparrow}}{R_{\uparrow\downarrow}+R_{\uparrow\uparrow}}\]
where $P$ is the polarization of the electron beam (20\%) and
$R_{\uparrow\uparrow}$ ($R_{\uparrow\downarrow}$) is the reflectivity
for electrons with spin parallel (anti-parallel) to the magnetization of the film. The white dashed lines at 2 ML film thickness in figure
\ref{fig:1}e and f clearly separate a uniform black region from a
modulated red and blue region. The featureless black region indicates
a perfect symmetry in both Fe and Co films reflectivity. This can be
explained by noting that the Curie temperature (T$_c$) of Fe/W(110)
films is above room temperature only for film thickness greater than
2 ML \cite{elmers:2031}. We attribute the absence of spin asymmetry
in thin Co films to a similar depression of the Curie temperature. The
remarkably perfect symmetry of the measurement above T$_c$ supports the
magnetic origin of the asymmetry measured below T$_c$.

\begin{figure} \includegraphics[width=8.5cm]{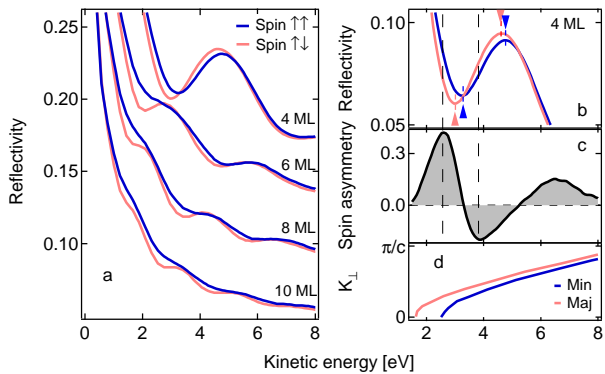}
\caption{\label{fig:2a}(Color online) (a) Energy dependence of the
reflectivity for spin parallel and anti-parallel to the Co/W(110)
film magnetic axis for different film thickness. The energy scans are
incrementally shifted by 0.05. (b) Expanded view of the spin dependence
of the reflectivity for the 4 ML thick film. (c) Spin asymmetry for a
4 ML thick Co/W(110) film obtained from fig 1b. (d) Band structure calculation for a 4 ML thick Co/W(110) film\cite{scheunemann:787}.} \end{figure}

Figure \ref{fig:2a}a shows energy scans of the spin dependent reflectivity
for different Co film thicknesses. For each film thickness, two
main observations can be made about the spin dependent reflectivity:
\begin{enumerate} \item{The reflectivity for the anti-parallel spin
(red curve) shows a red shift with respect to the case of parallel spin
(blue curve).} \item{The reflectivity oscillation in the case of spin
parallel (blue curve) are damped with respect to the anti-parallel
spin case (red curve).} \end{enumerate} These observations suggest that
two mechanisms contribute to the total spin asymmetry. The spectra for
film thickness bigger than 5-6 ML are mainly dominated by the damping
mechanism. We attribute the damping to different mean free paths for
spin parallel or anti-parallel to the majority spin. Indeed it is
known that the so-called universal curve predicts too large a value
of the inelastic mean free path (IMFP) for transition metals at low
energy, and spin dependent values of the IMFP corresponding to a few
monolayers have been reported at the energies considered in this paper
\cite{pappas:504,passek:103,getzlaff:467}. These results suggests that
SPLEEM can be used as a tool to extract information on the spin dependent
mean free path, a fundamental quantity for many properties of solids. A
more detailed study is required.

The red shift of the anti-parallel spin reflectivity spectra can be
clearly observed in figure \ref{fig:2a}b, where the spin dependent
reflectivity for a 4 ML thick Co/W(110) film is shown.  We see that
the large gradients of the reflectivity induced by the quantum well
states are slightly shifted for electrons of opposite spin due to the
exchange splitting of the band structure. The minima and the maxima of
the reflectivity indicated with the blue and red arrow are shifted by 0.3
eV and 0.15 eV respectively. It is the combination of this energy shift
with the large gradients in the reflectivity that causes significant
enhancement or reduction of the magnetic asymmetry as shown in figure
\ref{fig:2a}c. This seems to be the dominant mechanism at small film
thickness even in the presence of some inelastic damping. 

Figure \ref{fig:2a}d shows a band structure calculation along the perpendicular direction for a 4 ML Co/W(110) thin film \cite{scheunemann:787}. From the band structure calculation and the measured spin asymmetry, one can see that the first maxima at 2.6 eV of the spin asymmetry may have a contribution from the exchange-split band gap. But there is no band gap at the second maxima at 4.7 eV and one can see again the strong contribution of the finite size at small film thickness.

\begin{figure} \includegraphics[width=8.5cm]{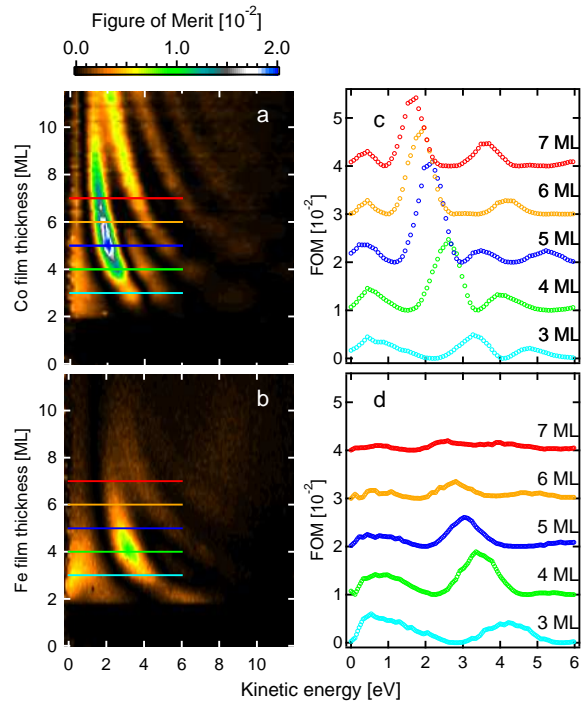}
\caption{\label{fig:2}(Color online) (a,b) Color-scale plots of the
``figure of merit'' for Co/W(110) and Fe/W(110) thin films respectively,
as a function of the electron kinetic energy (horizontal axis) and film
thickness (vertical axis). (c,d) The respective figure of merit, plotted
as a function of energy for Co and Fe at different thicknesses. For
clarity, the energy scans are incrementally shifted by 10$^{-2}$. The
location of the scans is shown by color horizontal lines in panel a and
b.} \end{figure}

In figures \ref{fig:2}a and b, the dependence of the figure of merit
(FOM) on electron kinetic energy (horizontal axis) and film thickness
(vertical axis) for Co/W(110) and Fe/W(110) is shown.  The figure of
merit combines the reflectivity and the spin asymmetry and is defined
by $A_s^2\frac{R_{\uparrow\downarrow}+R_{\uparrow\uparrow}}{2}$. This
whole map of the figure of merit emphasizes the important role of the QSE
oscillations for spin detection. The FOM is in fact very sensitive to the
film thickness and electron energy. A maximum for the FOM is reached only
in a narrow range of energy and film thickness. This feature must be taken
into account for further development of electron spin polarimetry. In
figure \ref{fig:2}c and d, several cuts in energy of the figure of merit
for Co/W(110) and Fe/W(110) are shown, respectively. The FOM of Co/W(110)
displays a pronounced maximum of 0.02 for 5 ML film thickness at an
electron kinetic energy of 2 eV. In Fe/W(110) we have found a maximal
FOM of 0.01 at 3.2 eV for 4 ML film thickness. These results clearly show
that the Co/W(110) thin film gives higher value of FOM than the Fe/W(110)
thin film when the films are grown at room temperature.

\begin{figure} \includegraphics[width=8.5cm]{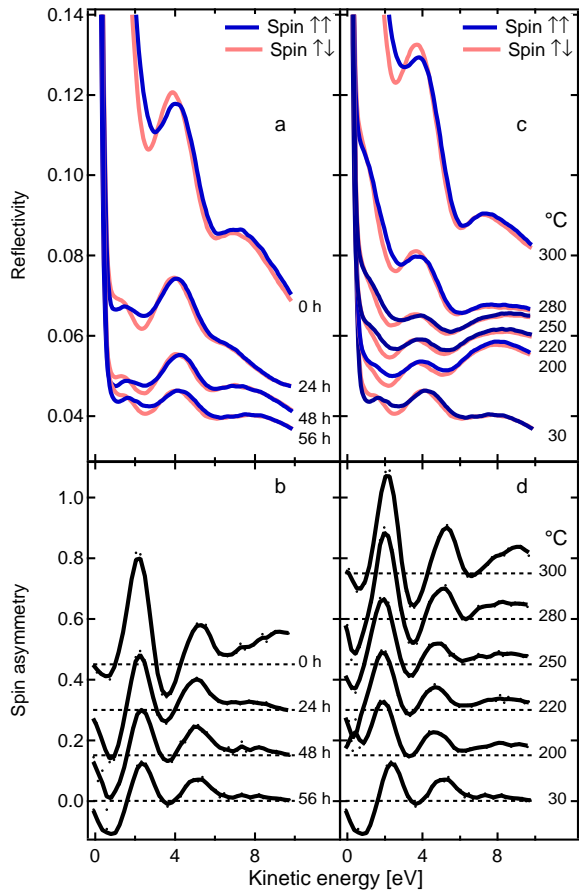}
\caption{\label{fig:3}(Color online) Aging effects and regeneration
for a 5 ML thick cobalt film: (a) Evolution of the reflectivity of
the film with time in UHV conditions. Red (blue) energy scans represent
incident electrons with spin polarization parallel (anti-parallel) to the
magnetization. For clarity, each curve is shifted by 0.01. (b) The spin
asymmetry calculated from the reflectivity distribution curves in panel
(a). The effect of annealing on the reflectivity and the corresponding
spin asymmetry are shown in (c) and (d), respectively, after 56 hours
of room-temperature aging in UHV.} \end{figure}

We also address a known difficulty in exchange-scattering polarimetry
related to surface conditions of the magnetic films used. Over the
course of many hours or days, adsorption of contamination from rest-gas
present in UHV considerably reduces both the reflectivity and the spin
asymmetry of the film (aging effects). In a spin-detector application
such aging effects cause degradation of the FOM over time and therefore
cause considerable losses of efficiency. To fully characterize the
Co/W(110) 5 ML thin film (the one leading to the highest FOM), we have
performed a detailed time and annealing temperature dependence of the
reflectivity. For this purpose, freshly grown films were stored in an
annex chamber with a base pressure of 2x10$^{-9}$ Torr to study the
effects of UHV contamination.

Figure \ref{fig:3}a and b shows how the high reflectivity and asymmetry
of a 5 ML thick Co film, measured every 24 hours for 3 days, evolve
with time. One can see that already after 24 hours the reflectivity is
substantially suppressed, the spin asymmetry is reduced by 50\%, and the
maximum FOM is diminished by almost 90\%. We found that this aging effect
can be easily reversed with a thermal treatment. Figures \ref{fig:3}c and
d show how careful annealing leads to substantial recovery of the film's
original exchange-scattering characteristics (75\% of the original FOM).
Additional heating (T$\geqslant$330$^{\circ}$ C.) cause the disruption
of the film with the formation of 3 dimensional islands surrounded by
a thin pseudomorphic film (Stranski-Krastanov film morphology). On the
other hand, the effect of UHV contamination on iron is even more severe,
and the FOM could not be significantly recovered by annealing before
film disruption. Neither film's FOM could be recovered after exposure to atmosphere.

A complete characterization would require the incident angular acceptance of both systems, but it cannot be measured by Low Energy Electron Microscopy where only backscattering of normal incident electrons is measured. Nevertheless the angular spread of electron is by definition very small in Angle Resolved Photoemission Spectroscopy, particularly with high efficiency electron spectrometers like time of flight spectrometers \cite{Hemmers:3809}.

\section{SUMMARY AND CONCLUSIONS}
\label{sec:SUMMARYANDCONCLUSIONS}
In conclusion, we presented a comparative analysis of the figure of merit
as a function of thickness, energy and time of two promising thin films
systems using a innovative technique based on spin polarized low energy
electron microscopy. We have shown that Co/W(110) thin film is an ideal
system for exchange-scattering based electron spin polarimetry, giving a
figure of merit as high as 0.02 with very low stray fields and thus without spurious asymmetry. In addition, because of the homogeneity of Co/W(110) thin films grown at room temperature and the possibility of reversing aging effects by moderate annealing, this system appears particularly stable for spin polarimetry applications.

\begin{acknowledgments} We thank G.-H. Gweon, M. Portalupi and Z. Q. Qiu
for helpful discussions. This work was supported by the Director,
Office of Science, Office of Basic Energy Sciences, Division of Materials
Sciences and Engineering, of the U.S. Department of Energy under Contract
No. DE-AC03-76SF00098. We acknowledge the support of the National Science
Foundation through Grant No. DMR-0349361, the A. P. Sloan Foundation and the Hellman Foundation.
\end{acknowledgments}

%\bibliography{jeffbiblio}

\begin{thebibliography}{21}
\expandafter\ifx\csname natexlab\endcsname\relax\def\natexlab#1{#1}\fi
\expandafter\ifx\csname bibnamefont\endcsname\relax
  \def\bibnamefont#1{#1}\fi
\expandafter\ifx\csname bibfnamefont\endcsname\relax
  \def\bibfnamefont#1{#1}\fi
\expandafter\ifx\csname citenamefont\endcsname\relax
  \def\citenamefont#1{#1}\fi
\expandafter\ifx\csname url\endcsname\relax
  \def\url#1{\texttt{#1}}\fi
\expandafter\ifx\csname urlprefix\endcsname\relax\def\urlprefix{URL }\fi
\providecommand{\bibinfo}[2]{#2}
\providecommand{\eprint}[2][]{\url{#2}}

\bibitem[{\citenamefont{Kessler}(1985)}]{kessler:1985}
\bibinfo{author}{\bibfnamefont{J.}~\bibnamefont{Kessler}},
  \emph{\bibinfo{title}{Polarized electrons}}
  (\bibinfo{publisher}{Springer-Verlag}, \bibinfo{address}{Berlin ; New York},
  \bibinfo{year}{1985}).

\bibitem[{\citenamefont{Pierce et~al.}(1988)\citenamefont{Pierce, Celotta,
  Kelley, and Unguris}}]{pierce:550}
\bibinfo{author}{\bibfnamefont{D.~T.} \bibnamefont{Pierce}},
  \bibinfo{author}{\bibfnamefont{R.~J.} \bibnamefont{Celotta}},
  \bibinfo{author}{\bibfnamefont{M.~H.} \bibnamefont{Kelley}},
  \bibnamefont{and} \bibinfo{author}{\bibfnamefont{J.}~\bibnamefont{Unguris}},
  \bibinfo{journal}{Nucl. Instr. and Meth. A} \textbf{\bibinfo{volume}{266}},
  \bibinfo{pages}{550} (\bibinfo{year}{1988}),
  \urlprefix\url{http://dx.doi.org/10.1016/0168-9002(88)90445-7}.

\bibitem[{\citenamefont{Fashold et~al.}(1992)\citenamefont{Fashold, Hammond,
  and Kirschner}}]{Fahsold:541}
\bibinfo{author}{\bibfnamefont{G.}~\bibnamefont{Fashold}},
  \bibinfo{author}{\bibfnamefont{M.~S.} \bibnamefont{Hammond}},
  \bibnamefont{and}
  \bibinfo{author}{\bibfnamefont{J.}~\bibnamefont{Kirschner}},
  \bibinfo{journal}{Solid State Commu.} \textbf{\bibinfo{volume}{84}},
  \bibinfo{pages}{541} (\bibinfo{year}{1992}),
  \urlprefix\url{http://dx.doi.org/10.1016/0038-1098(92)90186-D}.

\bibitem[{\citenamefont{Hillebrecht et~al.}(2002)\citenamefont{Hillebrecht,
  Jungblut, Wiebusch, Roth, Rose, Knabben, Bethke, Weber, Manderla, Rosowski
  et~al.}}]{hillbrecht:1229}
\bibinfo{author}{\bibfnamefont{F.~U.} \bibnamefont{Hillebrecht}},
  \bibinfo{author}{\bibfnamefont{R.~M.} \bibnamefont{Jungblut}},
  \bibinfo{author}{\bibfnamefont{L.}~\bibnamefont{Wiebusch}},
  \bibinfo{author}{\bibfnamefont{C.}~\bibnamefont{Roth}},
  \bibinfo{author}{\bibfnamefont{H.~B.} \bibnamefont{Rose}},
  \bibinfo{author}{\bibfnamefont{D.}~\bibnamefont{Knabben}},
  \bibinfo{author}{\bibfnamefont{C.}~\bibnamefont{Bethke}},
  \bibinfo{author}{\bibfnamefont{N.~B.} \bibnamefont{Weber}},
  \bibinfo{author}{\bibfnamefont{S.}~\bibnamefont{Manderla}},
  \bibinfo{author}{\bibfnamefont{U.}~\bibnamefont{Rosowski}},
  \bibnamefont{et~al.}, \bibinfo{journal}{Rev. Sci. Inst.}
  \textbf{\bibinfo{volume}{73}}, \bibinfo{pages}{1229} (\bibinfo{year}{2002}),
  \urlprefix\url{http://dx.doi.org/10.1063/1.1430547}.

\bibitem[{\citenamefont{Jungblut et~al.}(1992)\citenamefont{Jungblut, Roth,
  Hillebrecht, and Kisker}}]{jungblut:615}
\bibinfo{author}{\bibfnamefont{R.}~\bibnamefont{Jungblut}},
  \bibinfo{author}{\bibfnamefont{C.}~\bibnamefont{Roth}},
  \bibinfo{author}{\bibfnamefont{F.~U.} \bibnamefont{Hillebrecht}},
  \bibnamefont{and} \bibinfo{author}{\bibfnamefont{E.}~\bibnamefont{Kisker}},
  \bibinfo{journal}{Surf. Sci.} \textbf{\bibinfo{volume}{269-270}},
  \bibinfo{pages}{615} (\bibinfo{year}{1992}),
  \urlprefix\url{http://dx.doi.org/10.1016/0039-6028(92)91320-B}.

\bibitem[{\citenamefont{Hammond et~al.}(1992)\citenamefont{Hammond, Fahsold,
  and Kirschner}}]{Hammond:6131}
\bibinfo{author}{\bibfnamefont{M.}~\bibnamefont{Hammond}},
  \bibinfo{author}{\bibfnamefont{G.}~\bibnamefont{Fahsold}}, \bibnamefont{and}
  \bibinfo{author}{\bibfnamefont{J.}~\bibnamefont{Kirschner}},
  \bibinfo{journal}{Phys. Rev. B} \textbf{\bibinfo{volume}{45}},
  \bibinfo{pages}{6131} (\bibinfo{year}{1992}),
  \urlprefix\url{http://link.aps.org/abstract/PRB/v45/p6131}.

\bibitem[{\citenamefont{Bertacco and Ciccacci}(1999)}]{bertacco:265}
\bibinfo{author}{\bibfnamefont{R.}~\bibnamefont{Bertacco}} \bibnamefont{and}
  \bibinfo{author}{\bibfnamefont{F.}~\bibnamefont{Ciccacci}},
  \bibinfo{journal}{Surf. Sci.} \textbf{\bibinfo{volume}{419}},
  \bibinfo{pages}{265} (\bibinfo{year}{1999}),
  \urlprefix\url{http://dx.doi.org/10.1016/S0039-6028(98)00805-X}.

\bibitem[{\citenamefont{Bertacco et~al.}(1998)\citenamefont{Bertacco, Merano,
  and Ciccacci}}]{bertacco:2050}
\bibinfo{author}{\bibfnamefont{R.}~\bibnamefont{Bertacco}},
  \bibinfo{author}{\bibfnamefont{M.}~\bibnamefont{Merano}}, \bibnamefont{and}
  \bibinfo{author}{\bibfnamefont{F.}~\bibnamefont{Ciccacci}},
  \bibinfo{journal}{Applied Physics Letters} \textbf{\bibinfo{volume}{72}},
  \bibinfo{pages}{2050} (\bibinfo{year}{1998}),
  \urlprefix\url{http://link.aip.org/link/?APL/72/2050/1}.

\bibitem[{\citenamefont{Bertacco et~al.}(2002)\citenamefont{Bertacco, Marcon,
  Trezzi, Duo, and Ciccacci}}]{bertacco:3867}
\bibinfo{author}{\bibfnamefont{R.}~\bibnamefont{Bertacco}},
  \bibinfo{author}{\bibfnamefont{M.}~\bibnamefont{Marcon}},
  \bibinfo{author}{\bibfnamefont{G.}~\bibnamefont{Trezzi}},
  \bibinfo{author}{\bibfnamefont{L.}~\bibnamefont{Duo}}, \bibnamefont{and}
  \bibinfo{author}{\bibfnamefont{F.}~\bibnamefont{Ciccacci}},
  \bibinfo{journal}{Review of Scientific Instruments}
  \textbf{\bibinfo{volume}{73}}, \bibinfo{pages}{3867} (\bibinfo{year}{2002}),
  \urlprefix\url{http://link.aip.org/link/?RSI/73/3867/1}.

\bibitem[{\citenamefont{Zdyb and Bauer}(2002{\natexlab{a}})}]{zdyb:1485}
\bibinfo{author}{\bibfnamefont{R.}~\bibnamefont{Zdyb}} \bibnamefont{and}
  \bibinfo{author}{\bibfnamefont{E.}~\bibnamefont{Bauer}},
  \bibinfo{journal}{Surf. Rev. Lett.} \textbf{\bibinfo{volume}{9}},
  \bibinfo{pages}{1485} (\bibinfo{year}{2002}{\natexlab{a}}),
  \urlprefix\url{http://dx.doi.org/10.1142/S0218625X02003925}.

\bibitem[{\citenamefont{Zdyb and Bauer}(2002{\natexlab{b}})}]{zdyb:166403}
\bibinfo{author}{\bibfnamefont{R.}~\bibnamefont{Zdyb}} \bibnamefont{and}
  \bibinfo{author}{\bibfnamefont{E.}~\bibnamefont{Bauer}},
  \bibinfo{journal}{Physical Review Letters} \textbf{\bibinfo{volume}{88}},
  \bibinfo{eid}{166403} (pages~\bibinfo{numpages}{4})
  (\bibinfo{year}{2002}{\natexlab{b}}),
  \urlprefix\url{http://link.aps.org/abstract/PRL/v88/e166403}.

\bibitem[{\citenamefont{Bauer et~al.}(2002)\citenamefont{Bauer, Duden, and
  Zdyb}}]{bauer:2327}
\bibinfo{author}{\bibfnamefont{E.}~\bibnamefont{Bauer}},
  \bibinfo{author}{\bibfnamefont{T.}~\bibnamefont{Duden}}, \bibnamefont{and}
  \bibinfo{author}{\bibfnamefont{R.}~\bibnamefont{Zdyb}}, \bibinfo{journal}{J.
  Phys. D} \textbf{\bibinfo{volume}{35}}, \bibinfo{pages}{2327}
  (\bibinfo{year}{2002}),
  \urlprefix\url{http://dx.doi.org/10.1088/0022-3727/35/19/301}.

\bibitem[{\citenamefont{Gradmann and Waller}(1982)}]{gradmann:539}
\bibinfo{author}{\bibfnamefont{U.}~\bibnamefont{Gradmann}} \bibnamefont{and}
  \bibinfo{author}{\bibfnamefont{G.}~\bibnamefont{Waller}},
  \bibinfo{journal}{Surf. Sci.} \textbf{\bibinfo{volume}{116}},
  \bibinfo{pages}{539} (\bibinfo{year}{1982}),
  \urlprefix\url{http://dx.doi.org/10.1016/0039-6028(82)90363-6}.

\bibitem[{\citenamefont{Gardiner}(1983)}]{gardiner:213}
\bibinfo{author}{\bibfnamefont{T.~M.} \bibnamefont{Gardiner}},
  \bibinfo{journal}{Thin Solid Films} \textbf{\bibinfo{volume}{105}},
  \bibinfo{pages}{213} (\bibinfo{year}{1983}),
  \urlprefix\url{http://dx.doi.org/10.1016/0040-6090(83)90287-0}.

\bibitem[{\citenamefont{Bauer}(1999)}]{bauer:9365}
\bibinfo{author}{\bibfnamefont{E.}~\bibnamefont{Bauer}}, \bibinfo{journal}{J.
  Phys. C} \textbf{\bibinfo{volume}{11}}, \bibinfo{pages}{9365}
  (\bibinfo{year}{1999}),
  \urlprefix\url{http://dx.doi.org/10.1088/0953-8984/11/48/303}.

\bibitem[{\citenamefont{Scheunemann et~al.}(1997)\citenamefont{Scheunemann,
  Feder, Henk, Bauer, Duden, Pinkvos, Poppa, and Wurm}}]{scheunemann:787}
\bibinfo{author}{\bibfnamefont{T.}~\bibnamefont{Scheunemann}},
  \bibinfo{author}{\bibfnamefont{R.}~\bibnamefont{Feder}},
  \bibinfo{author}{\bibfnamefont{J.}~\bibnamefont{Henk}},
  \bibinfo{author}{\bibfnamefont{E.}~\bibnamefont{Bauer}},
  \bibinfo{author}{\bibfnamefont{T.}~\bibnamefont{Duden}},
  \bibinfo{author}{\bibfnamefont{H.}~\bibnamefont{Pinkvos}},
  \bibinfo{author}{\bibfnamefont{H.}~\bibnamefont{Poppa}}, \bibnamefont{and}
  \bibinfo{author}{\bibfnamefont{K.}~\bibnamefont{Wurm}},
  \bibinfo{journal}{Solid State Commu.} \textbf{\bibinfo{volume}{104}},
  \bibinfo{pages}{787} (\bibinfo{year}{1997}),
  \urlprefix\url{http://dx.doi.org/10.1016/S0038-1098(97)00357-8}.

\bibitem[{\citenamefont{Elmers et~al.}(1995)\citenamefont{Elmers, Hauschild,
  Fritzsche, Liu, Gradmann, and Kohler}}]{elmers:2031}
\bibinfo{author}{\bibfnamefont{H.~J.} \bibnamefont{Elmers}},
  \bibinfo{author}{\bibfnamefont{J.}~\bibnamefont{Hauschild}},
  \bibinfo{author}{\bibfnamefont{H.}~\bibnamefont{Fritzsche}},
  \bibinfo{author}{\bibfnamefont{G.}~\bibnamefont{Liu}},
  \bibinfo{author}{\bibfnamefont{U.}~\bibnamefont{Gradmann}}, \bibnamefont{and}
  \bibinfo{author}{\bibfnamefont{U.}~\bibnamefont{Kohler}},
  \bibinfo{journal}{Phys. Rev. Lett.} \textbf{\bibinfo{volume}{75}},
  \bibinfo{pages}{2031} (\bibinfo{year}{1995}),
  \urlprefix\url{http://link.aps.org/abstract/PRL/v75/p2031}.

\bibitem[{\citenamefont{Pappas et~al.}(1991)\citenamefont{Pappas, Kamper,
  Miller, Hopster, Fowler, Brundle, Luntz, and Shen}}]{pappas:504}
\bibinfo{author}{\bibfnamefont{D.~P.} \bibnamefont{Pappas}},
  \bibinfo{author}{\bibfnamefont{K.-P.} \bibnamefont{Kamper}},
  \bibinfo{author}{\bibfnamefont{B.~P.} \bibnamefont{Miller}},
  \bibinfo{author}{\bibfnamefont{H.}~\bibnamefont{Hopster}},
  \bibinfo{author}{\bibfnamefont{D.~E.} \bibnamefont{Fowler}},
  \bibinfo{author}{\bibfnamefont{C.~R.} \bibnamefont{Brundle}},
  \bibinfo{author}{\bibfnamefont{A.~C.} \bibnamefont{Luntz}}, \bibnamefont{and}
  \bibinfo{author}{\bibfnamefont{Z.-X.} \bibnamefont{Shen}},
  \bibinfo{journal}{Phys. Rev. Lett.} \textbf{\bibinfo{volume}{66}},
  \bibinfo{pages}{504} (\bibinfo{year}{1991}),
  \urlprefix\url{http://link.aps.org/abstract/PRL/v66/p504}.

\bibitem[{\citenamefont{Passek et~al.}(1996)\citenamefont{Passek, Donath, and
  Ertl}}]{passek:103}
\bibinfo{author}{\bibfnamefont{F.}~\bibnamefont{Passek}},
  \bibinfo{author}{\bibfnamefont{M.}~\bibnamefont{Donath}}, \bibnamefont{and}
  \bibinfo{author}{\bibfnamefont{K.}~\bibnamefont{Ertl}}, \bibinfo{journal}{J.
  Magn. Magn. Mater.} \textbf{\bibinfo{volume}{159}}, \bibinfo{pages}{103}
  (\bibinfo{year}{1996}),
  \urlprefix\url{http://dx.doi.org/10.1016/0304-8853(95)00940-X}.

\bibitem[{\citenamefont{Getzlaff et~al.}(1993)\citenamefont{Getzlaff, Bansmann,
  and Sch{\"o}nhense}}]{getzlaff:467}
\bibinfo{author}{\bibfnamefont{M.}~\bibnamefont{Getzlaff}},
  \bibinfo{author}{\bibfnamefont{J.}~\bibnamefont{Bansmann}}, \bibnamefont{and}
  \bibinfo{author}{\bibfnamefont{G.}~\bibnamefont{Sch{\"o}nhense}},
  \bibinfo{journal}{Solid State Commu.} \textbf{\bibinfo{volume}{87}},
  \bibinfo{pages}{467} (\bibinfo{year}{1993}),
  \urlprefix\url{http://dx.doi.org/10.1016/0038-1098(93)90799-S}.

\bibitem[{\citenamefont{Hemmers et~al.}(1998)\citenamefont{Hemmers, Whitfield,
  Glans, Wang, Lindle, Wehlitz, and Sellin}}]{Hemmers:3809}
\bibinfo{author}{\bibfnamefont{O.}~\bibnamefont{Hemmers}},
  \bibinfo{author}{\bibfnamefont{S.~B.} \bibnamefont{Whitfield}},
  \bibinfo{author}{\bibfnamefont{P.}~\bibnamefont{Glans}},
  \bibinfo{author}{\bibfnamefont{H.}~\bibnamefont{Wang}},
  \bibinfo{author}{\bibfnamefont{D.~W.} \bibnamefont{Lindle}},
  \bibinfo{author}{\bibfnamefont{R.}~\bibnamefont{Wehlitz}}, \bibnamefont{and}
  \bibinfo{author}{\bibfnamefont{I.~A.} \bibnamefont{Sellin}},
  \bibinfo{journal}{Review of Scientific Instruments}
  \textbf{\bibinfo{volume}{69}}, \bibinfo{pages}{3809} (\bibinfo{year}{1998}),
  \urlprefix\url{http://dx.doi.org/10.1063/1.1149183}.

\end{thebibliography}

\end{document}